\documentclass[aps,pra,a4paper,onecolumn,superscriptaddress,nofootinbib,longbibliography]{revtex4-2}

\usepackage{lmodern}
\usepackage{amsmath,amssymb,amsthm}
\usepackage{physics}
\usepackage{graphicx}
\usepackage{booktabs}
\usepackage{microtype}
\usepackage{tikz}
\usetikzlibrary{arrows.meta,shapes.geometric,positioning}
\usepackage[caption=false]{subfig}
\usepackage{algorithm}
\usepackage[noend]{algpseudocode}
\usepackage{enumitem}
\usepackage{multirow}
\usepackage{array}

\newcommand{\maybeincludegraphics}[2]{%
	\IfFileExists{#1}%
	{\includegraphics[width=#2]{#1}}%
	{\fbox{\parbox[c][0.25\textheight][c]{#2}%
			{\centering\ttfamily\small \detokenize{#1}\\[0.5em]\normalfont\small (file not found)}}}%
}

\theoremstyle{definition}

\newcommand{\nmax}{n_{\max}}
\newcommand{\nmodes}{n_{\mathrm{modes}}}
\newcommand{\nparams}{n_{\mathrm{params}}}

\newcommand{\Haf}{\mathrm{Haf}}
\newcommand{\WLN}{\mathrm{WLN}}

\usepackage[colorlinks=true, allcolors=blue, hyperfootnotes=false]{hyperref}
\hypersetup{
	colorlinks = true,
	linkcolor  = blue,
	citecolor  = blue,
	urlcolor   = blue,
	pdftitle   = {Rapid GBS Circuit Screening for GKP States  via a Two-Stage Machine Learning Surrogate},
	pdfauthor  = {Author Name(s)}
}

\begin{document}
	
	\title{Rapid Gaussian Boson Sampling Circuit Screening for GKP States Creation via a\texorpdfstring{\\}{ }%
		Two-Stage Machine Learning Surrogate}
	
	\author{Mohammad Amin Khanpour}
	\affiliation{Department of Physics, Shahid Beheshti University, Tehran, Iran}
	
	\author{Hossein Davoodi Yeganeh \footnote {corresponding author: h.yeganeh@ariaquanta.com, h.yeganeh@ssau.ac.ir}}
	\affiliation{AriaQuanta Quantum Computing Center, Tehran, Iran}
	\affiliation{Quantum Research Center, Shahid Sattari University of Aeronautical Sciences and Technology, Tehran, Iran}
	
	\begin{abstract}
		Gottesman--Kitaev--Preskill (GKP) states are essential non-Gaussian resources for fault-tolerant photonic quantum computing, enabling logical qubit encoding with intrinsic robustness against errors. Several approaches to GKP state preparation have been explored, including measurement-based protocols in circuit QED and trapped-ion systems, cat-state breeding, and photon-subtraction schemes. However, these methods are either restricted to specific platforms or require deep non-Gaussian resource chains with exponentially low success probabilities. Gaussian Boson Sampling (GBS) offers a compelling all-photonic alternative by generating non-Gaussian states through measurement-induced nonlinearity, without the need for matter-based ancilla or active feedforward. Nevertheless, its practical implementation is limited by the exponential computational cost of evaluating matrix hafnians---\#P-complete functions that govern photon-number probabilities. To address this challenge, we introduce a two-stage Histogram Gradient Boosting surrogate pipeline that predicts, without any hafnian computation, the optimal heralding pattern, circuit fidelity, and post-selection probability for candidate GBS circuits, while reserving exact quantum simulation exclusively for surrogate-selected candidates. Trained on  circuit configurations across 3--5 optical modes, the surrogate achieves 90.0\% GKP-detection accuracy on a held-out set, representing a 23.7 percentage-point improvement over the baseline, with a fidelity mean absolute error of 0.032 and a log-scale post-selection probability $R^2 = 0.837$, reducing the total simulation burden by approximately 90\%.

	\end{abstract}
	
	\maketitle
	
	\section{Introduction}
	\label{sec:introduction}
	
	Quantum computation promises solutions to classically intractable problems in cryptography, optimisation, and simulation. However, realising this potential requires hardware capable of performing high-fidelity operations at scale~\cite{r16}. Among the physical platforms currently under development, photonic quantum computing has emerged as a particularly promising candidate due to its room-temperature operation, low decoherence, and natural compatibility with quantum communication infrastructure~\cite{r17,r24}.	
	A central challenge in the development of practical photonic quantum processors is the generation of non-Gaussian resource states, among which Gottesman--Kitaev--Preskill (GKP) states play a central role~\cite{r1}. GKP states encode a logical qubit within a single bosonic mode in a way that is intrinsically resilient to small displacement errors in both quadratures. Moreover, they are fully compatible with Gaussian operations such as beamsplitters, squeezers, and phase rotations, which are naturally implemented in photonic hardware. When supplemented with GKP magic states, this framework enables universal fault-tolerant quantum computation~\cite{r2,r3,r32}, thereby making the preparation of GKP states a key requirement for scalable photonic quantum processors.
	The generation of GKP states remains a nontrivial task. Ideal GKP states require infinite squeezing and are therefore unphysical; in practice, finite-energy approximations must be prepared with sufficiently high fidelity, typically on the order of $F \ge 0.90$ with respect to the target finite-energy GKP state~\cite{r3,r12}, in order to provide meaningful protection against logical errors. 
	Several approaches to GKP state preparation have been investigated. Measurement-based schemes in circuit QED and trapped-ion platforms have shown promising performance~\cite{r18,r19,r29}, but they are inherently tied to matter-based architectures and are therefore not directly compatible with fully photonic implementations. Cat-state breeding and photon-subtraction techniques are fully optical and hardware-compatible; however, they typically require long sequences of non-Gaussian operations, which results in exponentially suppressed success probabilities~\cite{r20,r37}. Iterative protocols offer a more structured route toward target state generation, but they rely on real-time adaptive feedforward, introducing substantial experimental overhead and implementation complexity~\cite{r21}.
	In contrast, Gaussian boson sampling (GBS)~\cite{r22,r25,r26,r27} provides a fully photonic alternative. In this approach, a multimode Gaussian optical circuit—comprising squeezed-state sources, a linear interferometer, and photon-number-resolving (PNR) detectors—generates non-Gaussian states through measurement-induced nonlinearities, without the need for matter-based ancilla systems or active feedforward control.
	When specific heralded photon-number patterns are post-selected across all but one output mode, the remaining mode is projected into a non-Gaussian state that can closely approximate the target GKP wavefunction. The central engineering challenge is therefore to jointly determine the circuit parameters—namely squeezing amplitudes, interferometer angles and phases—and the appropriate heralding pattern, while simultaneously satisfying the dual constraints of high state fidelity and a sufficiently high post-selection probability.
	The evaluation of a single GBS circuit configuration is computationally expensive, as photon-number probabilities are expressed in terms of matrix hafnians~\cite{r6,r14}, which are \#P-complete functions that scale exponentially with photon number. For circuits with a Fock-space truncation of $n_{\max} = 12$ and five modes, a single evaluation requires approximately five minutes on a modern workstation. As a result, systematic exploration of the parameter space—requiring thousands of such evaluations—becomes computationally intractable without efficient approximation methods or surrogate models.
	This bottleneck motivates the use of surrogate modelling: a learned function that approximates the mapping from circuit parameters to performance metrics at a computational cost orders of magnitude lower than that of exact simulation~\cite{r31}. Such models enable rapid screening during parameter-space exploration, while restricting exact simulations to the final validation of the most promising candidates.
	This paper presents the \emph{GKP Circuit ML Pipeline v2}, a two-stage gradient-boosted surrogate framework designed to address the circuit screening problem in a systematic manner. At its core, a machine learning surrogate is a data-driven approximation of an expensive simulation function, trained on precomputed input--output pairs to learn the mapping from circuit parameters to performance metrics at a computational cost that is several orders of magnitude lower than that of the original simulator.
	The surrogate developed in this work—based on \texttt{HistGradientBoosting} models from \texttt{scikit-learn}—learns this mapping from 689 pre-optimised GBS circuit configurations, each annotated with ground-truth fidelity and post-selection probability obtained via exact quantum simulation. At inference time, it predicts the optimal heralding pattern, circuit fidelity, and post-selection probability for a new candidate circuit in approximately 1--5\,ms, without requiring any hafnian computations.	
	The pipeline architecture is guided by three design principles that distinguish it from a straightforward application of off-the-shelf regression methods. First, it adopts a cascade structure: a Stage~1 \texttt{HistGradientBoostingClassifier} predicts the optimal heralding pattern directly from circuit parameters, while Stage~2 regressors condition their predictions of fidelity and post-selection probability on this inferred pattern, thereby explicitly leveraging the physical dependence between herald photon-number structure and achievable output-state quality. 
	Second, it incorporates a physics-informed feature representation that augments the raw circuit parameters with eleven domain-derived aggregate statistics, capturing aspects such as total squeezing strength, interferometer coupling uniformity, and phase coherence. This representation allows a single fixed-dimensional model to generalise across circuits with varying mode counts without requiring topology-specific retraining. 
	Third, and most importantly for experimental robustness, the surrogate is not used in isolation: every circuit it identifies as GKP-capable ($F \ge 0.90$) is subsequently validated using full exact quantum simulation via Strawberry Fields~\cite{r5} and \texttt{thewalrus}~\cite{r14}, yielding confirmed values of fidelity, post-selection probability, Wigner function, and Wigner logarithmic negativity. This conditional validation strategy effectively turns the surrogate into a computational gating mechanism, which filters candidate circuits prior to expensive simulation rather than replacing it entirely. This distinction becomes particularly important when the model is applied outside its training distribution, as highlighted by the systematic failure cases discussed in Section~\ref{sec:casestudies}.
	The pipeline is trained on 689 pre-simulated circuit configurations spanning 3--5 optical modes, Fock-space truncations of $\nmax \in \{4,8,12\}$, and target squeezing levels $\Delta \in \{3\text{--}11\}\,\mathrm{dB}$. Evaluation is performed on a 170-sample holdout set and further stress-tested on nine representative circuits, including two deliberately constructed failure cases designed to probe the surrogate’s distributional limits. 
	The study also provides a candid characterization of the pipeline’s limitations, including a systematic failure mode associated with squeezing-parameter sign conventions outside the training distribution, as well as a cascade vulnerability stemming from the $36\%$ misclassification rate of the Stage~1 pattern predictor. 
	Section~\ref{sec:background} introduces the necessary background on GKP states, Gaussian boson sampling (GBS) circuit architectures, and the computational complexity of the hafnian. Section~\ref{sec:methods} details the pipeline design, dataset construction, feature engineering, and model architecture. Section~\ref{sec:evaluation} presents quantitative results, while Section~\ref{sec:casestudies} provides case-by-case comparisons between surrogate predictions and full quantum simulations. Finally, Section~\ref{sec:discussion} discusses limitations and potential mitigation strategies, and Section~\ref{sec:conclusion} concludes the paper.
	
	\section{Background}
	\label{sec:background}
	
	\subsection{GKP States}
	\label{subsec:gkp}
	
	GKP states encode a logical qubit in the infinite-dimensional Hilbert space of a single harmonic oscillator by distributing logical codewords over a periodic lattice in phase space~\cite{r1,r30}. The central physical idea is that logical information is stored in the \emph{periodicity} of the wavefunction rather than in any single degree of freedom, which makes the encoding naturally robust against small phase-space displacements. In particular, displacements smaller than $\sqrt{\pi}/2$ in either quadrature can be corrected via syndrome measurements without destroying the encoded logical state. 
	
	This structure differs fundamentally from discrete-variable error-correcting codes: the correctable error set is continuous rather than discrete, and protection arises from the geometric regularity of the phase-space lattice rather than from Hamming distance between codewords. To formalise this construction, the square-lattice encoding defines the logical computational basis states $\ket{\mu}_L$ for $\mu \in \{0,1\}$ as stabilised by the Weyl--Heisenberg displacement operators:
	\begin{equation}
		S_q = \exp\!\bigl(i\,2\sqrt{\pi}\,\hat{q}\bigr),
		\qquad
		S_p = \exp\!\bigl(-i\,2\sqrt{\pi}\,\hat{p}\bigr).
		\label{eq:stabilisers}
	\end{equation}
	These stabilisers enforce a periodic phase-space structure, rendering the code intrinsically robust to small displacement errors in both quadratures~\cite{r40,r36}.
	
	Since ideal GKP states require infinite squeezing and are therefore unphysical, practical implementations rely on finite-energy approximations characterised by an envelope parameter $\Delta$, which governs the trade-off between state fidelity and physical normalisability. The finite-energy logical codewords can be written as
	\begin{equation}
		\ket{\mu}_L \propto \hat{E}_L\,\ket{\mu}_{\mathrm{ideal}},
		\label{eq:finiteenergy}
	\end{equation}
	where $\hat{E}_L = \exp(-\Delta^2 \hat{n})$ denotes a Gaussian envelope operator acting on Fock space~\cite{r23}. In this formulation, larger values of $\Delta$ correspond to a tighter localisation of the state in phase space and a closer approximation to the ideal codeword, at the expense of an increased mean photon number. 
	
	The fidelity between a finite-energy approximation and the ideal codeword increases monotonically with $\Delta$,
	\begin{equation}
		F(\Delta) \xrightarrow{\Delta \to \infty} 1,
		\quad \text{monotonically.}
		\label{eq:fidelitymonotone}
	\end{equation}
	
	In the Fock basis, the logical state $\ket{0}_L$ has support only on even photon-number eigenstates:
	\begin{equation}
		\ket{0}_L = \sum_{k} c_{2k}\,\ket{2k},
		\label{eq:fock0}
	\end{equation}
	where the coefficients $c_{2k}$ satisfy the normalisation condition $\sum_k |c_{2k}|^2 = 1$. In practice, these coefficients are obtained via numerical diagonalisation of the finite-energy GKP Hamiltonian,
	\begin{equation}
		H_{\mathrm{GKP}} = -\cos\!\bigl(2\sqrt{\pi}\,\hat{q}\bigr)
		- \cos\!\bigl(2\sqrt{\pi}\,\hat{p}\bigr),
		\label{eq:gkpham}
	\end{equation}
	which enforces the underlying phase-space lattice structure~\cite{r23}. 
	
	Similarly, the logical state $\ket{1}_L$ is supported exclusively on odd Fock states:
	\begin{equation}
		\ket{1}_L = \sum_{k} c_{2k+1}\,\ket{2k+1}.
		\label{eq:fock1}
	\end{equation}
	
	This parity structure—restricting $\ket{0}_L$ to even photon numbers and $\ket{1}_L$ to odd photon numbers—has direct implications for circuit design, since valid heralding patterns must respect this symmetry. The fidelity between a prepared state $\ket{\psi}$ and a target finite-energy codeword $\ket{\mu}_L$ is defined as
	\begin{equation}
		F = \bigl|\,{}_L\!\braket{\mu}{\psi}\bigr|^2.
		\label{eq:fidelity}
	\end{equation}
	A fidelity threshold of $F \ge 0.90$ is commonly regarded as necessary—though not sufficient—for practical fault-tolerant operation~\cite{r12}. 
	
	State quality is further quantified using the Wigner logarithmic negativity (WLN):
	\begin{equation}
		\mathrm{WLN} = \log\!\left(\iint \bigl|W(q,p)\bigr|\,dq\,dp\right).
		\label{eq:wln}
	\end{equation}
	A positive WLN ($\mathrm{WLN} > 0$) is a necessary, though not sufficient, condition for quantum computational advantage beyond what is classically simulable using Gaussian methods~\cite{r11}. Importantly, a state may satisfy $F \ge 0.90$ relative to a GKP target while still exhibiting $\mathrm{WLN} \approx 0$, in which case it remains classically simulable and provides no meaningful advantage for fault-tolerant quantum computation.
	
	\subsection{Gaussian Boson Sampling and GKP State Generation}
	\label{subsec:gbs}
	
	Gaussian boson sampling (GBS) provides an all-photonic mechanism for generating non-Gaussian states via measurement-induced nonlinearity, without the need for matter-based ancilla systems or active feedforward~\cite{r22}. The central idea is as follows: a multimode Gaussian state—prepared using single-mode squeezers and evolved through a passive linear interferometer—is measured with photon-number-resolving (PNR) detectors on all but one output mode. 
	
	By post-selecting on a specific heralding pattern $\mathbf{m} = (m_1, m_2, \ldots, m_{\nmodes - 1})$ across the measured modes, the remaining unmeasured mode is projected into a conditional non-Gaussian state that is fully determined by the circuit parameters and the chosen herald outcome. By appropriately tuning the squeezing amplitudes and interferometer angles, this conditional state can be engineered to approximate a target GKP wavefunction with high fidelity. 
	
	The probability of observing a given herald pattern, denoted $p(\mathbf{m})$, determines the rate at which usable resource states are produced. This introduces an intrinsic trade-off between state fidelity and preparation efficiency, which motivates the surrogate-based optimisation framework developed in this work.
	
	Formally, Gaussian boson sampling (GBS) can be described as follows. A multimode Gaussian state, characterised by its covariance matrix $\Sigma$ and displacement vector $\boldsymbol{\mu}$, is transformed through a passive linear optical circuit and subsequently measured. The resulting photon-number statistics are determined by hafnians of submatrices of the transformed covariance matrix, rendering exact classical simulation widely believed to be computationally intractable~\cite{r6,r25,r39}. 
	
	In the context of state preparation, the post-selection probability associated with a herald pattern $\mathbf{m}$ in an ideal lossless circuit is given by
	\begin{equation}
		p(\mathbf{m}) \propto
		\frac{\bigl|\Haf(A_{\mathbf{m}})\bigr|^2}{\displaystyle\prod_i m_i!},
		\label{eq:pselection}
	\end{equation}
	where $A_{\mathbf{m}}$ denotes the submatrix of the covariance structure selected by the pattern and the underlying circuit parameters. 
	
	In this setting, valid herald patterns are those satisfying $\sum_i m_i = \nmax$, with each entry being a non-negative integer, $m_i \ge 0$, across the $\nmodes - 1$ measured modes.
	
	\subsection{The Hafnian and Computational Complexity}
	\label{subsec:hafnian}
	The hafnian of a $2n \times 2n$ symmetric matrix $A$ is defined as
	\begin{equation}
		\Haf(A)
		= \sum_{\sigma \in \mathrm{PMP}(2n)}
		\prod_{(i,j)\in\sigma} A_{ij},
		\label{eq:hafnian}
	\end{equation}
	where the sum runs over all perfect matchings of $2n$ elements. This quantity is \#P-hard to evaluate~\cite{r6}, and is therefore at least as computationally difficult as counting the solutions to an NP problem. The best known classical algorithms scale as $\mathcal{O}(2^n \cdot n^2)$, which restricts exact evaluation to relatively small system sizes. In practice, the \texttt{thewalrus} library implements highly optimised C++ routines with symmetry-based accelerations, although the underlying exponential scaling remains unchanged.
	
	The practical implications for GKP circuit design are significant. For five-mode circuits with $\nmax = 12$, a single evaluation of a (circuit configuration, herald pattern) pair requires approximately five minutes on a modern workstation using optimised routines from the \texttt{thewalrus} library~\cite{r14}. Given up to 15 valid herald patterns in this setting, exhaustive evaluation of a single circuit therefore requires roughly 75 minutes. As a result, screening $1{,}000$ candidate configurations would amount to approximately $1{,}250$ CPU-hours, making brute-force exploration impractical for interactive design or large-scale optimisation. Even for smaller instances such as $\nmax = 8$ with four modes, individual evaluations still require on the order of 10–30 seconds, which remains prohibitive for extensive search.
	
	The surrogate pipeline circumvents this bottleneck by reducing screening time to milliseconds per circuit, while restricting expensive quantum simulations to those candidates that are predicted to be GKP-capable.
	
	\subsection{GBS Circuit Architecture}
	\label{subsec:circuit}
	
	The GBS circuit studied in this work follows the architecture established in~\cite{r4}
	and is implemented using the Strawberry Fields quantum photonics library~\cite{r5}.
	It consists of three stages.
	
	\textbf{Stage~1 -- Single-mode squeezing.}
	Each of the $\nmodes$ input vacuum modes is prepared in a squeezed state using
	$\mathrm{Sgate}(r_k,0)$, where $r_k$ is the squeezing amplitude (related to squeezing
	in dB by $r = \ln(10)\times\mathrm{dB}/20$).
	Squeezing amplitudes are optimised within $r_k \in [-r_{\max}, r_{\max}]$, where
	$r_{\max}$ corresponds to 15\,dB.
	
	\textbf{Stage~2 -- Linear interferometer.}
	A rectangular Clements-decomposition interferometer~\cite{r13} transforms the squeezed
	product state into a multimode entangled Gaussian state.
	It consists of $\nmodes(\nmodes-1)/2$ beamsplitter gates $\mathrm{BSgate}(\theta_k,\phi_k)$,
	where $\theta_k \in [-\pi,\pi]$ is the mixing angle and $\phi_k \in [-\pi,\pi]$ is the
	relative phase.
	The Clements architecture guarantees universality: any passive linear-optical unitary
	$U \in U(\nmodes)$ can be realised by appropriate choice of these parameters.
	
	\textbf{Stage~3 -- Photon-number-resolving measurement and post-selection.}
	The output modes of the interferometer are directed to photon-number-resolving (PNR)
	detectors.
	Post-selection is applied to the first $\nmodes - 1$ output modes: only measurement
	outcomes matching the target herald pattern
	$\mathbf{m} = (m_1, m_2, \ldots, m_{\nmodes-1})$ are accepted.
	Conditioned on this outcome, the remaining undetected output mode is projected into a
	non-Gaussian heralded state whose structure is entirely determined by the circuit
	parameters and the specific pattern.
	The post-selection probability $p(\mathbf{m})$, governed by the hafnian of the relevant
	covariance submatrix~(Eq.~\eqref{eq:pselection}), sets the rate at which usable
	resource states are produced.
	All simulations in this work are performed in the lossless idealisation ($T_k = 1$ at
	all points in the circuit), so the heralded output is a pure state; the code
	infrastructure supports optional loss channels at multiple circuit locations, but these
	are not employed in the results reported here.
	
	The total number of free circuit parameters is:
	\begin{equation}
		\nparams = \nmodes + \nmodes(\nmodes - 1),
		\label{eq:nparams}
	\end{equation}
	giving 9, 16, and 25 parameters for three-, four-, and five-mode circuits respectively.
	
	\begin{figure}[h!]
		\centering
		\maybeincludegraphics{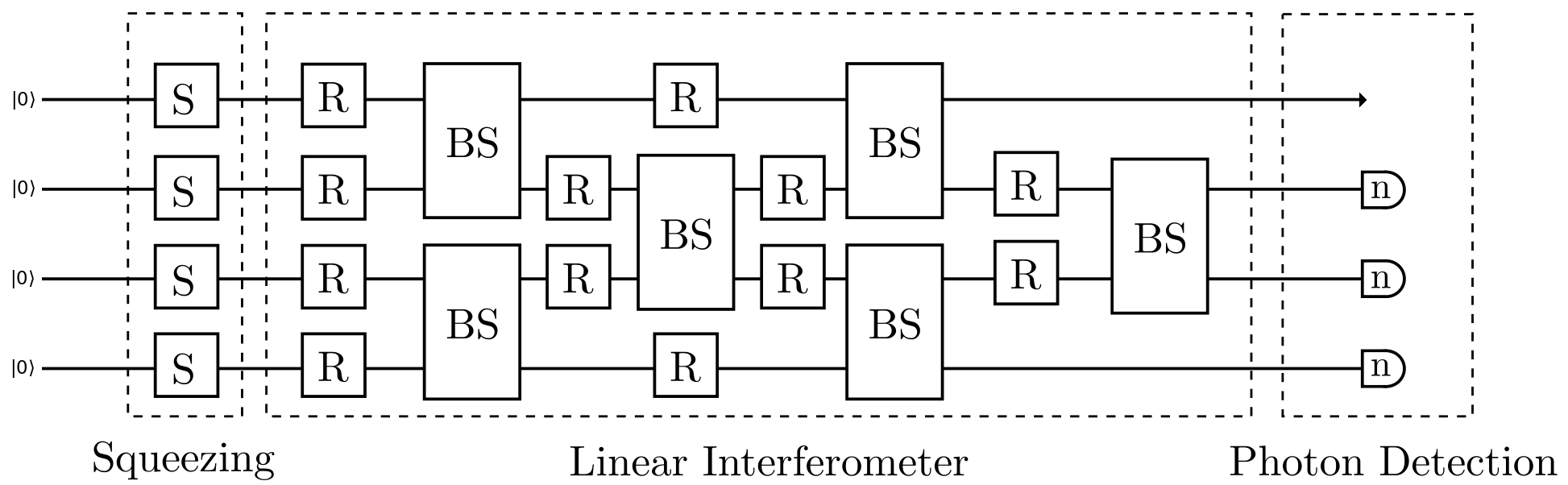}{0.95\textwidth}
		\caption{%
			GBS circuit schematic for GKP state preparation.
			The circuit consists of (left to right): single-mode squeezing gates on all
			$\nmodes$ input modes, a rectangular Clements beamsplitter interferometer,
			photon-number-resolving (PNR) detectors
			post-selecting on herald pattern
			$\mathbf{m} = (m_1,\ldots,m_{\nmodes-1})$ on the first $\nmodes-1$ modes, and
			the undetected output mode which collapses into the heralded GKP state.
		}
		\label{fig:circuit}
	\end{figure}
	\section{Methods}
	\label{sec:methods}
	
	\subsection{Pipeline Overview}
	\label{subsec:overview}
	
	The GKP Circuit ML Pipeline v2 is distinguished from a naive surrogate approach by four structural features that reflect the underlying physics of the problem.
	First, it employs a \emph{cascade architecture}: Stage~1 predicts the optimal heralding pattern directly from circuit parameters, while Stage~2 conditions its predictions of fidelity and post-selection probability on the Stage~1 output. This design explicitly leverages the dependence of achievable fidelity on the specific photon-number configuration being post-selected, reformulating the task as conditionally structured rather than fully joint multivariate regression.
	Second, the feature representation augments raw circuit parameters with eleven physics-inspired summary statistics that encode prior knowledge about squeezing distributions, interferometer coupling structure, and phase coherence. This enables generalisation across circuits of varying size within a unified fixed-dimensional representation.
	Third, and most importantly from an experimental standpoint, the surrogate is not used as a standalone predictor. Any circuit classified as GKP-capable ($F \ge 0.90$) is subsequently validated via full quantum simulation using Strawberry Fields~\cite{r5} and \texttt{thewalrus}~\cite{r14}, producing verified estimates of fidelity, success probability, Wigner function, and Wigner logarithmic negativity. In this way, the surrogate acts as a computational gating mechanism rather than a replacement for simulation.
	Fourth, the study provides a detailed characterisation of failure modes, including a systematic error arising from squeezing-parameter sign conventions outside the training distribution, as well as a cascade-level vulnerability associated with the $36\%$ misclassification rate of the Stage~1 classifier.	
	\begin{figure}[h!]
		\centering
		\maybeincludegraphics{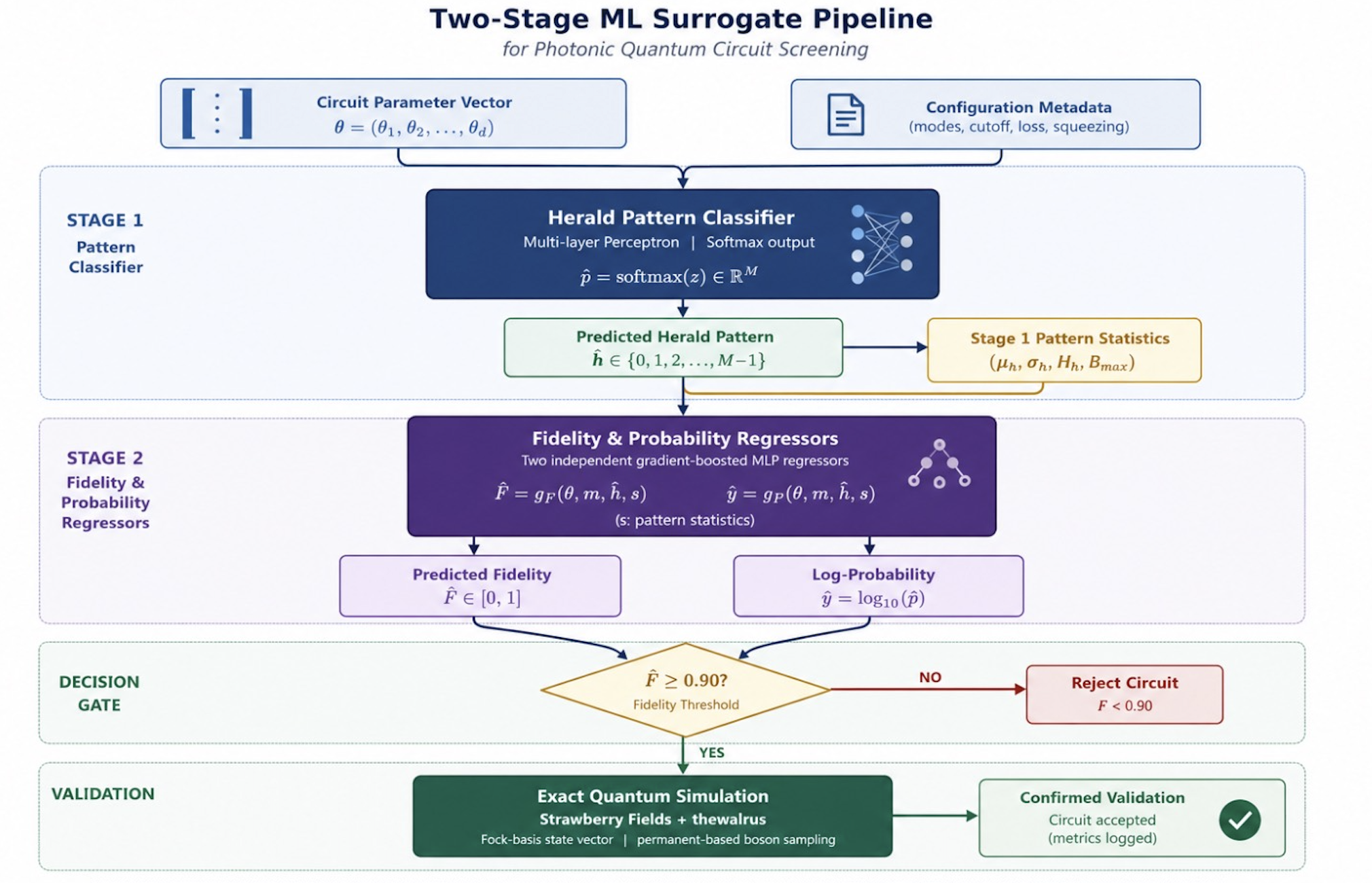}{0.90\textwidth}
		\vspace{-10pt}
		\caption{
			Two-stage ML surrogate pipeline block diagram.
			Stage~1 (Pattern Classifier) takes the circuit parameter vector and configuration
			metadata as input and outputs the predicted herald pattern.
			Stage~2 (Fidelity and Probability Regressors) takes the same input augmented with
			the Stage~1 pattern statistics and outputs predicted fidelity $F$ and
			log-probability $y = \log_{10}(p)$.
			If $F \ge 0.90$, the circuit is passed to exact quantum simulation
			(Strawberry Fields + \texttt{thewalrus}) for confirmed validation.
		}
		\label{fig:pipeline}
	\end{figure}
	
	\subsection{Dataset Construction}
	\label{subsec:dataset}
	
	The training dataset was assembled from 423 pickle files generated by running the GKP
	preparation circuit optimiser derived from the Xanadu approximate-GKP-prep
	repository~\cite{r4,r23}.
	Each pickle file encodes the optimised circuit parameters and corresponding fidelity
	and probability values for a specific $(\nmax, \nmodes, \mathbf{m}, \Delta)$
	configuration.
	Circuit configurations span $\nmodes \in \{3,4,5\}$ optical modes,
	$\nmax \in \{4,8,12\}$ Fock-space truncation, and
	$\Delta \in \{3\text{--}11\}\,\mathrm{dB}$ target envelope squeezing.
	
	For each configuration, circuit parameters were optimised in two sequential phases.
	Phase~1 applied the SLSQP gradient-based minimiser~\cite{r33} from a randomly sampled starting
	point to minimise:
	\begin{equation}
		\mathcal{L}(\boldsymbol{\theta})
		= -\bigl[F(\boldsymbol{\theta}) + p(\boldsymbol{\theta})\bigr],
		\label{eq:lossphase1}
	\end{equation}
	subject to bounds $r_k \in [-r_{\max}, r_{\max}]$ for squeezing amplitudes and
	$\theta_k, \phi_k \in [-\pi,\pi]$ for all interferometer angles.
	This cost function equally weights fidelity maximisation and probability maximisation.
	Phase~2 applied the basin-hopping global optimiser~\cite{r34} with 50 random jumps and SLSQP as
	the local minimiser, using a modified cost:
	\begin{equation}
		\mathcal{L}'(\boldsymbol{\theta})
		= -\bigl[F(\boldsymbol{\theta}) + 0.1\,p(\boldsymbol{\theta})\bigr].
		\label{eq:lossphase2}
	\end{equation}
	The reduced weight on probability in the global phase deliberately deprioritises
	probability during random exploration, preventing premature convergence into basins
	characterised by high post-selection probability but low fidelity.
	Both the Phase~1 and Phase~2 solutions were retained as separate training samples,
	yielding two data points per pickle file.
	From 423 files, this produced 846 samples.
	An additional 13 benchmark configurations from prior GKP simulation studies
	(\texttt{paper\_data}) were included, yielding 859 total samples.
	
	The dataset was partitioned as follows: the 846 \texttt{test\_data} samples were
	randomly shuffled (random seed $= 42$) and split 80/20 into 676 training and 170 test
	samples.
	All 13 paper benchmark samples were added to the training partition, giving 689 training
	samples and 170 holdout samples.
	
	Of the 689 training samples, 457 (66.3\%) have fidelity $F \ge 0.90$ and are labelled
	GKP-capable.
	This 2:1 imbalance was addressed through explicit sample weighting during model
	training: paper benchmark circuits received a weight multiplier of $10\times$, anchoring
	the model on precisely characterised reference configurations; GKP-capable samples
	received an additional multiplier of $2\times$, improving sensitivity at and near the
	$F = 0.90$ classification boundary.
	All weights were normalised to unit mean before training.
	
	\subsection{Feature Engineering}
	\label{subsec:features}
	
	A central design choice in the proposed pipeline is the use of a fixed-dimensional feature representation that accommodates circuits of varying size ($\nmodes \in \{3,4,5\}$) without requiring separate models for each topology. This is achieved by zero-padding all parameter vectors to match the maximum dimensions in the training set ($\max \nmodes = 5$, $\max n_{\mathrm{splitters}} = 10$) and concatenating four categories of descriptors.
	
	The first category consists of raw circuit parameters. The $\nmodes$ squeezing amplitudes $r_k$, the $\nmodes(\nmodes-1)/2$ beamsplitter mixing
	angles $\theta_k$, and the beamsplitter phases $\phi_k$ are each zero-padded to lengths
	5, 10, and 10 respectively, producing a 25-element raw parameter block.
	These features retain full information about the circuit, including the sign of
	squeezing amplitudes---a property that introduces a systematic failure mode when test
	circuits use sign conventions underrepresented in training
	(see Section~\ref{sec:casestudies}). The second category consists of configuration metadata. The scalar values $\nmax$, $\nmodes$, and $\Delta$ are appended directly.
	These three scalars are among the most physically informative features: the achievable
	fidelity ceiling is fundamentally constrained by $\nmax$ (which determines the accuracy
	of the Fock-basis GKP approximation) and $\Delta$ (which sets the target squeezing
	level). The third category consists of physics-derived aggregate features. Eleven statistics are computed from the raw circuit parameters to encode domain knowledge
	not directly recoverable from the zero-padded raw features, as defined in
	Table~\ref{tab:features}.
	
	\begin{table}[h]
		\centering
		\caption{Physics-derived aggregate features (Category~3).}
		\label{tab:features}
		\begin{tabular}{lll}
			\toprule
			\textbf{Feature Name} & \textbf{Definition} & \textbf{Physical Interpretation} \\
			\midrule
			\texttt{total\_sq\_power}  & $\sum_k r_k^2$
			& Total optical squeezing energy injected \\
			\texttt{max\_sq}           & $\max_k |r_k|$
			& Strongest single-mode squeezing \\
			\texttt{sq\_asymmetry}     & $\max(r_k) - \min(r_k)$
			& Spread of squeezing across modes \\
			\texttt{n\_negative\_sq}   & $\#\{k : r_k < 0\}$
			& Number of phase-displaced squeezers \\
			\texttt{mean\_sq\_abs}     & $\langle|r_k|\rangle$
			& Average squeezing per mode \\
			\texttt{mean\_coupling}    & $\langle\theta_k\rangle$
			& Average interferometer mixing strength \\
			\texttt{coupling\_spread}  & $\sigma(\theta_k)$
			& Uniformity of beamsplitter coupling \\
			\texttt{phase\_coherence}  & $\sigma(\phi_k)$
			& Spread of interferometer phases \\
			\texttt{near\_pi\_count}   & $\#\bigl\{k : \bigl||\phi_k| - \pi\bigr| < 0.1\bigr\}$
			& Number of near-$\pi$ phase rotations \\
			\texttt{nmax\_per\_mode}   & $\nmax / \nmodes$
			& Effective photons per mode \\
			\texttt{delta\_over\_nmax} & $\Delta / \nmax$
			& Squeezing quality per Fock level \\
			\bottomrule
		\end{tabular}
	\end{table}
	
	Note that \texttt{total\_sq\_power}, \texttt{max\_sq}, \texttt{mean\_sq\_abs}, and
	\texttt{sq\_asymmetry} are sign-invariant in magnitude.
	However, the raw squeezing parameters in Category~1 are sign-sensitive.
	This asymmetry---partially sign-invariant physics features combined with sign-sensitive
	raw parameters---is the root cause of the failure mode documented in
	Section~\ref{sec:casestudies}.
	Note also that \texttt{near\_pi\_count} counts phases close to $\pm\pi$ in absolute
	value (i.e.\ $||\phi_k| - \pi| < 0.1$), capturing the physical near-equivalence of
	$+\pi$ and $-\pi$ phases for the beamsplitter gate.
	
	The final category consists of Herald pattern statistics (Stage~2 only). Eight statistics summarising the herald pattern structure are appended to the Stage~2
	feature vector:
	$\texttt{pat\_sum} = \sum_i m_i$ (always equals $\nmax$),
	$\texttt{pat\_len} = \nmodes - 1$,
	and $\texttt{pat\_max}$, $\texttt{pat\_min}$, $\texttt{pat\_mean}$,
	$\texttt{pat\_std}$, $\texttt{pat\_median}$, $\texttt{pat\_range}$.
	These allow Stage~2 to condition its predictions on the photon-count distribution being
	post-selected.
	
	All features in both Stage~1 and Stage~2 vectors were standardised to zero mean and unit
	variance using \texttt{sklearn} \texttt{StandardScaler}~\cite{r28} fitted exclusively on the
	training set.
	Two independent scalers were maintained---one for the Stage~1 feature vector
	(Categories~1--3, dimension 39) and one for the Stage~2 feature vector
	(Categories~1--4, dimension 47)---preventing data leakage between splits and
	inconsistency in normalisation between pipeline stages.
	
	\subsection{Model Architecture and Cascade Design Rationale}
	\label{subsec:architecture}
	
	The three prediction targets---optimal herald pattern, circuit fidelity, and
	post-selection probability---are not independent.
	The fidelity and probability achievable by a circuit depend on which herald pattern is
	applied: different patterns probe different photon-number correlations in the multimode
	state, and the pattern that maximises fidelity is not always the one that maximises
	probability.
	The cascade design propagates the Stage~1 pattern prediction as a feature into Stage~2,
	allowing the fidelity and probability regressors to exploit the physical relationship
	between pattern geometry and circuit performance.
	
	\textbf{Stage~1: Herald Pattern Classifier.}
	A \texttt{HistGradientBoostingClassifier}~\cite{r28,r38} is trained on the Stage~1 feature vector
	(39 features) to predict the optimal herald pattern as a multiclass label.
	Herald patterns are represented as ampersand-delimited strings (e.g.,~\texttt{"3\&5"}
	for pattern $[3,5]$) and mapped to integer class labels via \texttt{LabelEncoder}.
	For circuit configurations whose predicted pattern falls outside the set of
	mathematically valid patterns for the given $(\nmax, \nmodes)$, a nearest-neighbour
	correction maps the prediction to the closest valid pattern by Euclidean distance.
	Hyperparameters for the multiclass case (more than two valid patterns):
	\texttt{max\_iter} $= 500$, \texttt{max\_depth} $= 8$, \texttt{learning\_rate} $= 0.05$,
	\texttt{min\_samples\_leaf} $= 3$, \texttt{l2\_regularization} $= 0.1$.
	For the binary case (exactly two valid patterns): \texttt{max\_iter} $= 300$,
	\texttt{max\_depth} $= 6$, \texttt{learning\_rate} $= 0.05$.
	
	\textbf{Stage~2a: Fidelity Regressor.}
	A \texttt{HistGradientBoostingRegressor} is trained on the Stage~2 feature vector
	(47 features, where Category~4 contains the statistics of the Stage~1 predicted
	pattern) to predict circuit fidelity $F \in [0,1]$.
	Predictions are clipped to $[0,1]$ at inference.
	The GKP-capability threshold $F \ge 0.90$ applied to the Stage~2a output is the primary
	classification gate.
	Hyperparameters: \texttt{max\_iter} $= 500$, \texttt{max\_depth} $= 8$,
	\texttt{learning\_rate} $= 0.05$, \texttt{min\_samples\_leaf} $= 5$,
	\texttt{l2\_regularization} $= 0.1$.
	
	\textbf{Stage~2b: Probability Regressor.}
	A second \texttt{HistGradientBoostingRegressor} predicts the log-transformed
	post-selection probability:
	\begin{equation}
		y = \log_{10}(p + \varepsilon), \qquad \varepsilon = 10^{-30},
		\label{eq:logtransform}
	\end{equation}
	where the offset $\varepsilon$ prevents undefined logarithms for configurations with
	numerically zero post-selection probability.
	Predictions are back-transformed at inference via $p = 10^y$ and clipped to $[0,1]$.
	The log transformation is essential: the raw probability spans approximately ten orders
	of magnitude across the training set (from $\sim\!10^{-12}$ to $\sim\!10^{-2}$), and
	regressing on untransformed values would cause the MSE loss to be dominated by the small
	number of high-probability configurations.
	In $\log_{10}$ scale, an MAE of $1.0$ corresponds to exactly one order of magnitude
	error---a natural and interpretable unit given the observed dynamic range.
	Hyperparameters: identical to Stage~2a.
	
	\subsection{Conditional Validation Architecture}
	\label{subsec:validation}
	
	The surrogate pipeline does not make final predictions autonomously.
	Instead, it implements a conditional validation structure in which exact quantum
	simulation is triggered for every circuit that passes the surrogate screening gate
	(predicted $F \ge 0.90$).
	Specifically:
	\begin{enumerate}[label=(\roman*)]
		\item The full GBS circuit is simulated using the Strawberry Fields Gaussian backend,
		which propagates the circuit covariance matrix $\Sigma$ and displacement vector
		$\boldsymbol{\mu}$ analytically through all squeezing and interferometer stages.
		\item The post-selection probability is computed using
		\texttt{density\_matrix\_element} from \texttt{thewalrus}, which evaluates the
		hafnian of the relevant covariance submatrix.
		\item The heralded state is extracted using \texttt{state\_vector} (lossless) or
		\texttt{density\_matrix} (lossy) from \texttt{thewalrus}.
		\item The fidelity between the heralded state and the target GKP Fock-basis
		representation is computed exactly.
		\item The Wigner function is evaluated on a $100 \times 100$ phase-space grid over
		$q, p \in [-7, 7]$ using an iterative algorithm adapted from QuTiP, and Wigner
		log-negativity is computed on a separate $200 \times 200$ grid by 2D numerical
		integration using \texttt{scipy.integrate.simpson}.
		\item Simulation results (fidelity, probability, Wigner function, WLN) are returned
		alongside the surrogate predictions and presented side by side.
	\end{enumerate}
	
	This architecture means the surrogate serves exclusively as a computational filter, not
	as the final authority on circuit performance.
	False-positive surrogate predictions are exposed immediately by the subsequent
	simulation.
	False-negative predictions---GKP-capable circuits rejected by the surrogate---are the
	true cost of the screening step, representing circuits never simulated and thus never
	discovered.
	For large circuits ($\nmax = 12$, five modes) where each (circuit, pattern) evaluation
	costs approximately five minutes and up to 15 valid patterns exist, exhaustive screening
	of $10{,}000$ candidates would require approximately $12{,}500$ CPU-hours.
	Restricting full simulation to only the $10\%$ of surrogate-endorsed candidates reduces
	this to approximately $1{,}250$ CPU-hours---a tenfold reduction---with the surrogate's
	pattern recommendation eliminating the need to evaluate all 15 patterns per approved
	circuit.
	
	\section{Quantitative Evaluation}
	\label{sec:evaluation}
	
	Any honest assessment of a binary classifier requires comparison against trivial
	baselines.
	For the GKP-detection task, two natural baselines exist.
	The majority-class baseline---always predicting ``GKP-capable''---achieves accuracy
	equal to the GKP-capable fraction: $457/689 = 66.3\%$ on the training set,
	approximately $66.3\%$ on the holdout.
	The random baseline---predicting ``GKP-capable'' with probability equal to the training
	class frequency---achieves approximately $66.3^2 + 33.7^2 \approx 55\%$.
	
	The surrogate pipeline achieves $90.0\%$ GKP-detection accuracy on the holdout set,
	representing a $23.7$ percentage-point improvement over the majority-class
	baseline---a meaningful and non-trivial result.
	However, this must be read in context: $10\%$ of holdout circuits are incorrectly
	classified, of which some fraction are false positives (GKP-incapable circuits passed
	to simulation) and the remainder are false negatives (GKP-capable circuits rejected).
	
	For the pattern classifier, the $64.0\%$ cross-validated accuracy represents a
	meaningful improvement over chance (which ranges from $6.7\%$ for 15-class
	configurations to $50\%$ for 2-class configurations).
	However, the $36\%$ error rate in Stage~1 propagates into Stage~2 feature errors with
	consequences for fidelity prediction quality that are discussed further in
	Section~\ref{sec:discussion}.
	
	Table~\ref{tab:baselines} contextualises the surrogate's GKP-detection accuracy
	against trivial baselines on the holdout set.
	Table~\ref{tab:metrics} presents all quantitative performance metrics across
	training, cross-validation (5-fold), and holdout evaluation.
	
	\begin{table}[h]
		\centering
		\caption{%
			GKP-detection accuracy of the surrogate pipeline against trivial baselines
			on the held-out test set ($n = 170$).
			The majority-class baseline always predicts ``GKP-capable'' and achieves
			the training-set class frequency ($457/689 = 66.3\%$); the random baseline
			predicts ``GKP-capable'' with that same probability, yielding
			$66.3^{2} + 33.7^{2} \approx 55\%$.
			Wilson-score 95\,\% confidence intervals are reported for the surrogate
			only, since the baselines are deterministic thresholds rather than
			stochastic predictors evaluated on the holdout.
			pp\;=\;percentage points.%
		}
		\label{tab:baselines}
		\begin{tabular}{@{}lccc@{}}
			\toprule
			\textbf{Method}
			& \textbf{Accuracy}
			& \textbf{95\,\% CI}
			& $\boldsymbol{\Delta}$\,\textbf{vs.\ majority} \\
			\midrule
			Random baseline
			& ${\approx}55.0\%$
			& ---
			& $-11.3$\,pp \\
			Majority-class baseline
			& $66.3\%$
			& ---
			& --- \\
			\midrule
			\textbf{Surrogate (ours)}
			& $\mathbf{90.0\%}$
			& $[84.8\%,\;93.6\%]$
			& $\mathbf{+23.7}$\,pp \\
			\bottomrule
		\end{tabular}
	\end{table}
	
	\begin{table}[h]
		\centering
		\caption{Complete quantitative performance metrics across all pipeline stages.}
		\label{tab:metrics}
		\begin{tabular}{llccc}
			\toprule
			\textbf{Model} & \textbf{Metric} & \textbf{CV Score}
			& \textbf{CV $\pm 2\sigma$} & \textbf{Holdout} \\
			\midrule
			Pattern Classifier        & Accuracy           & 0.640 & $\pm 0.067$ & ---   \\
			Fidelity Regressor        & $R^2$              & 0.689 & $\pm 0.294$ & 0.760 \\
			Fidelity Regressor        & MAE                & ---   & ---         & 0.032 \\
			Probability Regressor     & $R^2$ ($\log_{10}$)& 0.811 & $\pm 0.138$ & 0.837 \\
			Probability Regressor     & MAE ($\log_{10}$)  & ---   & ---         & 0.432 \\
			\midrule
			\textbf{GKP Detection}    & \textbf{Accuracy}  & ---   & ---         & \textbf{0.900} \\
			Majority-class baseline   & Accuracy           & ---   & ---         & $\approx 0.663$ \\
			\textbf{Improvement}      & $\Delta$ Accuracy  & ---   & ---         & $\mathbf{+0.237}$ \\
			\bottomrule
		\end{tabular}
	\end{table}
	
	\textbf{Fidelity regressor performance.}
	The holdout $R^2$ of $0.760$ indicates the model explains approximately $76\%$ of
	fidelity variance across the 170 holdout configurations.
	The MAE of $0.032$ is the more practically useful metric: fidelity predictions deviate
	from exact simulation by an average of $3.2$ percentage points.
	This is acceptable for circuits well above or well below $F = 0.90$, but is imprecise
	for circuits near the threshold.
	
	The cross-validation $R^2$ of $0.689 \pm 0.294$ warrants serious attention.
	The $2\sigma$ spread means individual fold $R^2$ values may range from approximately
	$0.40$ to $0.98$.
	A fold $R^2$ of $0.40$ implies that in some subsets of the parameter space, the model
	explains less than half the fidelity variance---approaching the performance of a simple
	mean predictor.
	
	\textbf{Probability regressor performance.}
	The holdout log-scale $R^2$ of $0.837$ and MAE of $0.432$ indicate the model predicts
	post-selection probabilities to within approximately half an order of magnitude on
	average.
	Given that probabilities span ten orders of magnitude across the dataset, this is a
	useful level of accuracy for experimental resource planning---reliable discrimination
	between circuits requiring minutes versus hours of data collection---but should not be
	used for precise rate estimation in experiments requiring narrow probability windows.
	
	\textbf{Paper benchmark results.}
	The 13 paper benchmark circuits achieve $R^2 = 1.00$ and MAE $\approx 0.000$ for both
	fidelity and probability.
	This result must be interpreted cautiously: these circuits received a $10\times$ training
	weight, effectively forcing the model to reproduce their values through local
	interpolation.
	This is evidence of memorisation of heavily-weighted training points, not of strong
	out-of-distribution generalisation.
	
	\textbf{Cascade vulnerability.}
	The $36\%$ Stage~1 pattern misclassification rate means Stage~2 receives incorrect
	pattern statistics as input in approximately one third of cases.
	The aggregate holdout MAE of $0.032$ mixes correctly-conditioned Stage~2 predictions
	($\approx 64\%$ of cases) with incorrectly-conditioned ones ($\approx 36\%$), potentially
	obscuring much larger errors in the misclassified subset.
	
	\begin{figure}[t]
		\centering
		\maybeincludegraphics{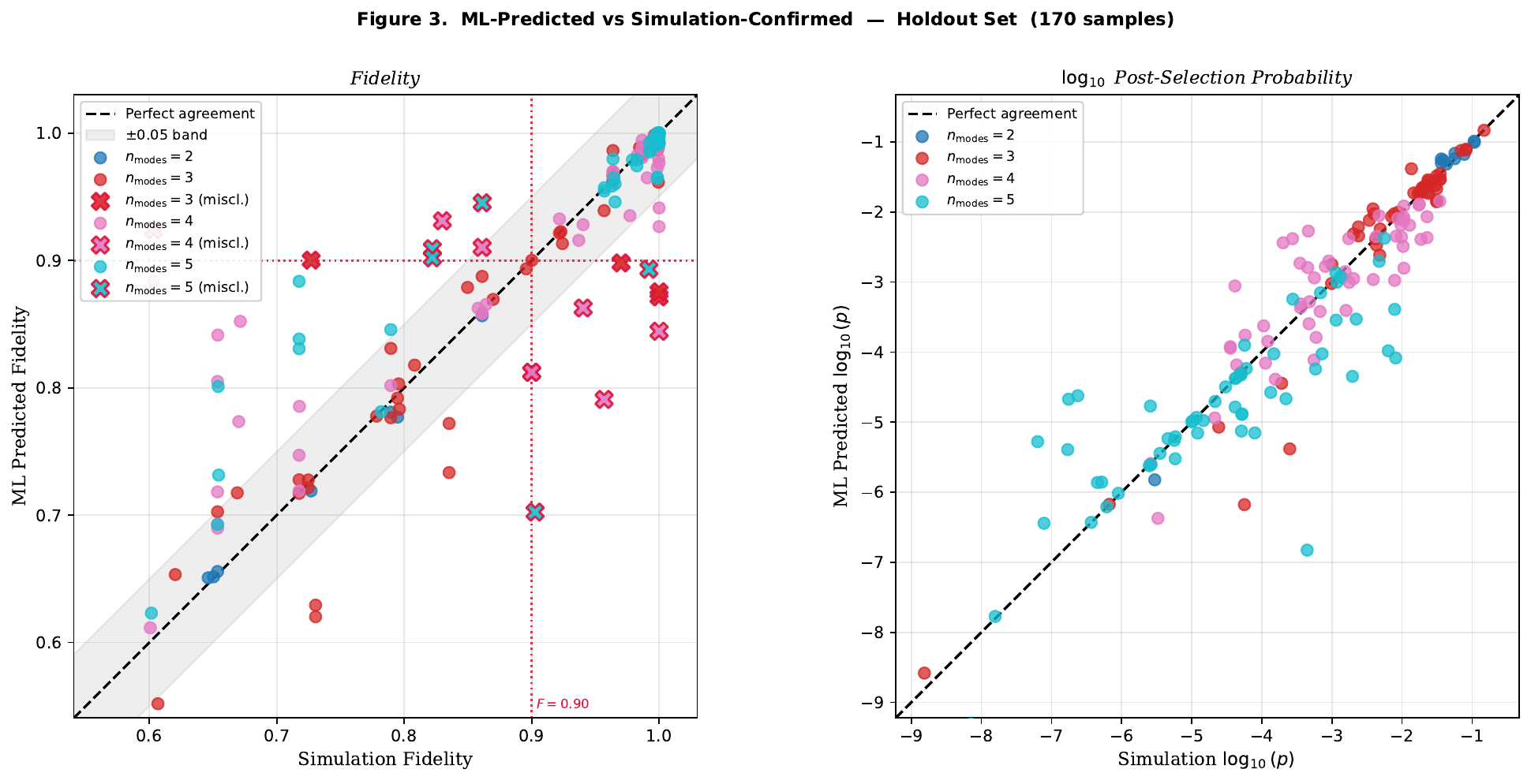}{0.95\textwidth}
		\caption{%
			Scatter plot of ML-predicted vs.\ simulation-confirmed fidelity (left) and
			$\log_{10}$ post-selection probability (right) on the 170-sample holdout set.
			Points are coloured by circuit mode count ($\nmodes \in \{3,4,5\}$).
			The dashed diagonal line indicates perfect agreement.
			The shaded band indicates $\pm 0.05$ fidelity error.
			Points above or below the $F = 0.90$ threshold line that are misclassified by the
			surrogate are highlighted.
		}
		\label{fig:scatter}
	\end{figure}
	
	\section{Case Studies: Surrogate Prediction vs.\ Exact Simulation}
	\label{sec:casestudies}
	
	Nine circuit configurations were selected to characterise the pipeline's end-to-end
	behaviour across a range of mode counts, Fock truncations, and squeezing targets.
	These circuits were deliberately chosen---not randomly sampled---to span diverse
	configurations: the paper demo circuit (C1), configurations at both ends of the $\nmax$
	range (C6 at $\nmax = 4$; F1 at $\nmax = 12$), and configurations designed to explore
	the sign-sensitivity failure mode (C3 vs.~F1; C4 vs.~F2).
	
	For each circuit, the pipeline executed the following sequence: Stage~1 predicted the
	optimal herald pattern; Stage~2 predicted fidelity and probability conditional on the
	Stage~1 pattern; if $F \ge 0.90$, exact Strawberry Fields simulation was performed; the
	Wigner function was computed and Wigner log-negativity evaluated numerically; surrogate
	predictions and simulation results were compared.
	
	Table~\ref{tab:casestudies} summarises all nine circuits.
	Circuits C1--C7 are successful predictions; F1 and F2 are systematic failure cases.
	
	\begin{table*}[t]
		\centering
		\caption{%
			End-to-end case study results.
			$\checkmark$ indicates correct pattern recommendation.
			Rows F1 and F2 are failure cases.
			$\Delta F = |F_{\mathrm{ML}} - F_{\mathrm{sim}}|$ is the absolute fidelity
			prediction error.
		}
		\label{tab:casestudies}
		\begin{tabular}{lcccccccc}
			\toprule
			\textbf{Config} & $\nmax/\nmodes/\Delta\,(\mathrm{dB})$
			& \textbf{Pattern} & $F_{\mathrm{ML}}$ & $F_{\mathrm{sim}}$
			& $\Delta F$ & $p_{\mathrm{ML}}$ & $p_{\mathrm{sim}}$ & WLN \\
			\midrule
			C1 (demo)  & $8/3/10$      & $[4,4]$ $\checkmark$     & 0.969 & 0.999 & 0.030
			& $4.22\times10^{-3}$ & $4.10\times10^{-3}$ & 0.700 \\
			C2         & $8/4/5$       & $[2,3,3]$ $\checkmark$   & 0.971 & 0.971 & 0.001
			& $9.20\times10^{-13}$& $8.67\times10^{-13}$& 0.115 \\
			C3         & $12/5/11\,(-)$& $[1,2,3,6]$ $\checkmark$ & 1.000 & 0.999 & 0.000
			& $1.89\times10^{-5}$ & $1.89\times10^{-5}$ & 0.736 \\
			C4         & $12/4/10\,(+)$& $[3,4,5]$ $\checkmark$   & 0.956 & 0.980 & 0.025
			& $8.95\times10^{-7}$ & $8.96\times10^{-7}$ & 0.700 \\
			C5         & $8/4/10$      & $[2,2,4]$ $\checkmark$   & 1.000 & 1.000 & 0.000
			& $1.35\times10^{-3}$ & $1.36\times10^{-3}$ & 0.696 \\
			C6         & $4/3/11$      & $[2,2]$ $\checkmark$     & 0.999 & 0.999 & 0.001
			& $1.57\times10^{-2}$ & $1.54\times10^{-2}$ & 0.640 \\
			C7         & $8/3/5$       & $[3,5]$ $\checkmark$     & 0.964 & 0.965 & 0.002
			& $3.30\times10^{-2}$ & $3.50\times10^{-2}$ & 0.022 \\
			\midrule
			F1 $\times$& $12/5/11\,(+)$& $[1,2,3,6]$ $\checkmark$ & 0.995 & 0.307 & 0.688
			& $1.36\times10^{-5}$ & $7.81\times10^{-5}$ & 0.518 \\
			F2 $\times$& $12/4/10\,(-)$& $[3,4,5]$ $\checkmark$   & 0.929 & 0.001 & 0.928
			& $2.50\times10^{-6}$ & $1.93\times10^{-5}$ & 0.563 \\
			\bottomrule
		\end{tabular}
	\end{table*}
	
	\textbf{Pattern recommendation.}
	The surrogate correctly recommended the herald pattern in all seven successful cases.
	This is the most operationally important correct prediction: an incorrect pattern
	recommendation not only conditions Stage~2 poorly but directs the subsequent simulation
	to the wrong post-selection, guaranteeing a wrong fidelity result even if exact
	simulation is performed.
	
	\textbf{Fidelity agreement.}
	Fidelity prediction errors range from essentially zero (C2: $\Delta F = 0.001$;
	C3: $\Delta F \approx 0$; C5: $\Delta F \approx 0$) to a maximum of
	$\Delta F = 0.030$ for C1 (the paper demo circuit).
	This 3 percentage-point underestimation does not constitute a classification error---both
	values are well above $0.90$---but illustrates that the surrogate can be systematically
	conservative for high-fidelity circuits.
	
	\textbf{Probability agreement.}
	Post-selection probability predictions span many orders of magnitude: from
	$9.20 \times 10^{-13}$ for C2 to $3.30 \times 10^{-2}$ for C7.
	Despite this range, surrogate and simulation agree closely in most cases---C3 achieves
	near-exact agreement and C4 achieves agreement to four significant figures.
	
	\textbf{Wigner log-negativity and the fidelity-only threshold problem.}
	The WLN values reveal a significant limitation of $F \ge 0.90$ as the sole quality
	criterion.
	C2 ($\WLN = 0.115$) and C7 ($\WLN = 0.022$) both satisfy the fidelity threshold but
	have WLN values approaching zero---a state with $\WLN \approx 0$ is essentially
	classically simulable and cannot provide quantum computational advantage for non-Clifford
	gate synthesis or magic state distillation~\cite{r11}.
	This is not a failure of the surrogate, which makes no WLN predictions; it is a
	limitation of the threshold criterion itself.
	The WLN pattern is physically interpretable: WLN correlates with both $\Delta$ and
	$\nmax$.
	Lower $\Delta$ (e.g., $5\,\mathrm{dB}$ for C7) produces a less squeezed finite-energy
	approximation with weaker negativity and lower WLN.
	Higher $\Delta$ (e.g., $10$--$11\,\mathrm{dB}$ for C1, C4, C5) produces sharper
	phase-space peaks, stronger negativity, and WLN consistently in the range $0.63$--$0.74$.
	
	\textbf{Systematic failure cases F1 and F2.}
	The surrogate predicts $F = 0.995$ for F1 and $F = 0.929$ for F2.
	Exact simulation reveals $F_{\mathrm{sim}} = 0.307$ for F1 and
	$F_{\mathrm{sim}} = 0.001$ for F2---errors of $\Delta F = 0.688$ and $\Delta F = 0.928$
	respectively.
	Without the mandatory simulation validation step, these circuits would be incorrectly
	endorsed as high-quality GKP state generators.
	The source of this failure is identifiable through controlled comparison:
	
	F1 uses squeezing parameters $[{+}1.701, -0.254, -0.711, -1.724, -1.725]$
	(predominantly positive first parameter) with configuration
	$(\nmax = 12, \nmodes = 5, \Delta = 11\,\mathrm{dB}, \mathbf{m} = [1,2,3,6])$.
	C3 uses squeezing parameters $[-1.701, -0.254, -0.711, -1.724, -1.725]$---the same
	magnitudes, negated first parameter---with identical configuration.
	F1 yields $F_{\mathrm{sim}} = 0.307$; C3 yields $F_{\mathrm{sim}} = 0.999$.
	The surrogate predicts $F = 0.995$ for F1 and $F = 0.9995$ for C3---near-identical
	surrogate predictions for circuits whose actual fidelities differ by more than $0.69$.
	The same pattern holds for F2 vs.~C4.
	
	This is not a random error.
	It is a systematic consequence of the training data's sign distribution: the optimised
	configurations apparently cluster more heavily in one sign convention for five-mode and
	four-mode circuits with high $\nmax$.
	When a test circuit falls in the underrepresented sign convention, the surrogate maps it
	to a high-fidelity training neighbourhood that does not correspond to its actual physical
	behaviour.
	The conditional simulation step prevents these failures from propagating: both F1 and F2
	trigger exact simulation, which immediately reveals the true fidelities.
	The surrogate fails; the pipeline does not.
	
	\begin{figure*}[t]
		\centering
		\maybeincludegraphics{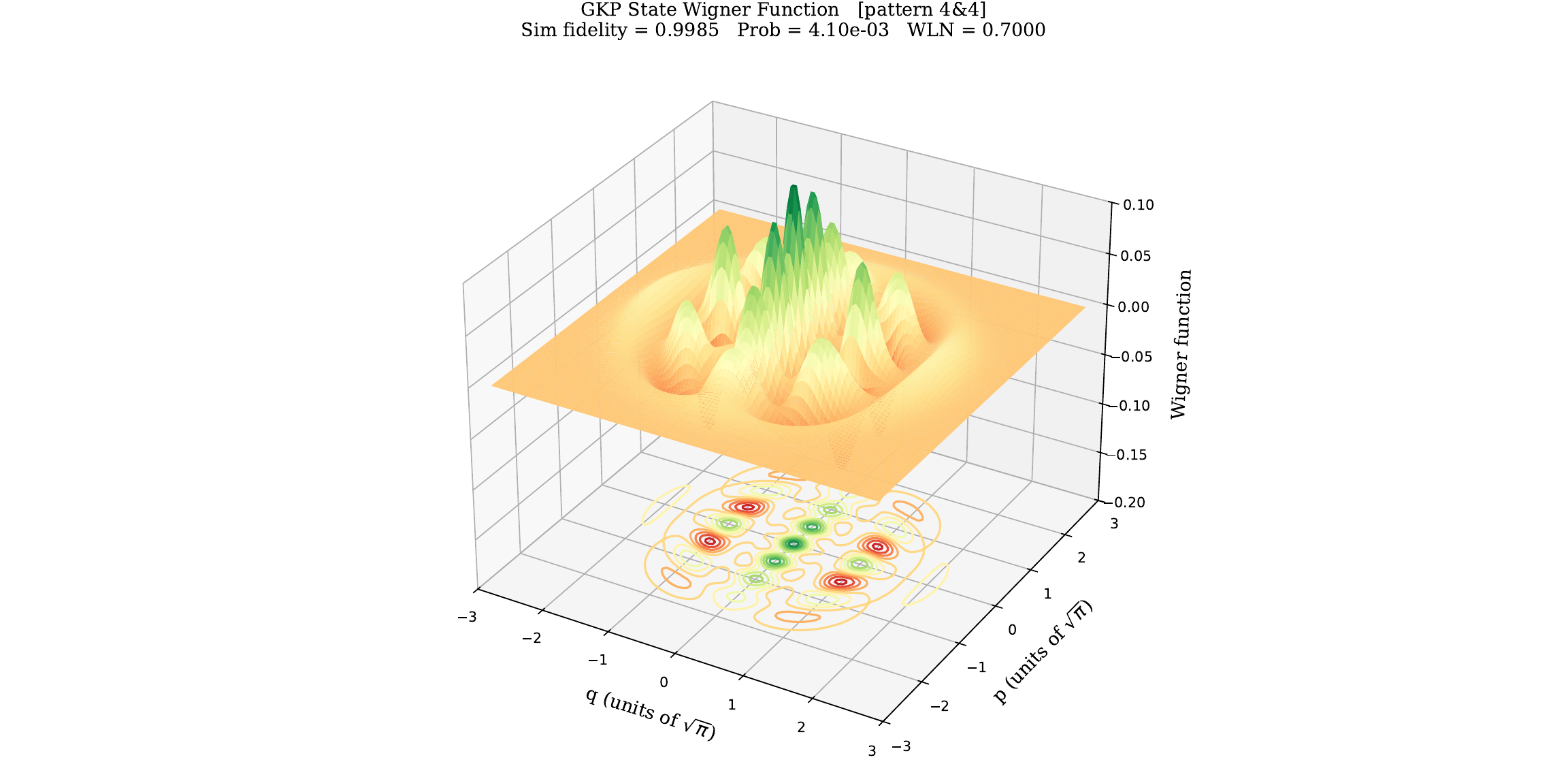}{0.95\textwidth}
		\caption{%
			Representative Wigner functions $W(q,p)$ from the exact simulation step, plotted
			over $q, p \in [-3,3]$ (in units of $\sqrt{\pi}$).
			\textbf{(a)}~Circuit C1 ($\nmax = 8$, three modes, $\Delta = 10\,\mathrm{dB}$,
			$\WLN = 0.700$): characteristic GKP lattice structure with prominent negativity
			(red/blue regions) at phase-space lattice points.
			\textbf{(b)}~Circuit C7 ($\nmax = 8$, three modes, $\Delta = 5\,\mathrm{dB}$,
			$\WLN = 0.022$): the GKP pattern is visually present but with negligible
			negativity, consistent with near-classical behaviour.
			\textbf{(c)}~Circuit F1 ($\nmax = 12$, five modes, $\Delta = 11\,\mathrm{dB}$,
			sign $+$): the Wigner function bears no resemblance to a GKP state, consistent
			with $F_{\mathrm{sim}} = 0.307$, confirming the catastrophic surrogate failure.
		}
		\label{fig:wigner}
	\end{figure*}
	
	\section{Discussion}
	\label{sec:discussion}
	
	For circuit configurations within the training distribution—specifically 3--5 mode GBS circuits with $\nmodes \in \{3,4,5\}$, $\nmax \in \{4,8,12\}$, and $\Delta \in \{3\text{--}11\}\,\mathrm{dB}$—the proposed surrogate achieves reliable performance in identifying GKP-capable circuits, reaching an accuracy of approximately $90\%$, corresponding to a 23.7 percentage-point improvement over the baseline. Within successfully validated cases, it correctly identifies suitable heralding patterns and predicts circuit fidelity with a mean absolute error of $0.032$. These outputs are produced in the millisecond regime, and any circuit that satisfies the GKP threshold is immediately subjected to full exact quantum simulation for verification.
	
	However, the surrogate does not generalise robustly beyond the training distribution. This limitation is primarily manifested through a sign-sensitivity failure mode, where deviations in squeezing-parameter conventions lead to degraded predictions. In addition, the model does not directly predict Wigner logarithmic negativity (WLN), and its fidelity regressor exhibits non-uniform performance across the parameter space, with cross-validated $R^2$ values decreasing to approximately $0.40$ in certain regimes. Furthermore, the Stage~1 pattern classifier, with an accuracy of around $64\%$, introduces cascading errors that propagate into Stage~2 predictions and are not fully captured by aggregate metrics.
	
	A key aspect of the framework is that it does not function as a standalone predictor. Instead, it operates as a computational filter: every circuit predicted to be GKP-capable is subsequently validated via exact quantum simulation, ensuring that reported fidelity, success probability, and Wigner-logarithmic-negativity values are physically grounded and independent of surrogate error. This separation between fast inference and exact verification is essential for maintaining reliability in practice.
	
	Several limitations remain. First, the sign-sensitivity issue can be mitigated through symmetry-augmented training data, sign-invariant feature construction, or explicit input canonicalisation to enforce consistent parameter conventions. Second, the observed cross-validation variability ($R^2 = 0.689 \pm 0.294$) suggests heterogeneous model reliability, which could be improved through active learning strategies~\cite{r35} that prioritise simulation in high-uncertainty regions. Third, the cascade vulnerability introduced by Stage~1 misclassification can be addressed either by improving training coverage in high-degeneracy regimes or by removing hard conditioning and allowing Stage~2 to model all valid herald patterns jointly. Finally, the absence of WLN prediction limits the ability to directly screen for non-classical computational utility; incorporating WLN as an additional regression target would extend the model beyond fidelity estimation.
	
	From a computational perspective, the advantage of the surrogate is substantial. For five-mode circuits with $\nmax = 12$, each circuit-pattern evaluation requires approximately five minutes using exact methods, and with up to 15 valid herald patterns this results in roughly 75 minutes per circuit. By contrast, the surrogate produces predictions in approximately 1--5 ms, yielding speedups on the order of $10^5$ for large configurations. In large-scale screening scenarios~\cite{r24}, where approximately 90\% of candidates are rejected prior to simulation, the total computational cost for evaluating $10{,}000$ circuits is reduced from approximately 12,500 CPU-hours under exhaustive evaluation to about 1,250 CPU-hours under surrogate-guided filtering.
	
	Finally, the reliability of the framework is strictly bounded by its training distribution. It is trained on $\nmodes \in \{3,4,5\}$, $\nmax \in \{4,8,12\}$, and $\Delta \in \{3\text{--}11\}\,\mathrm{dB}$, and performance degrades outside these ranges, including for different mode counts, higher Fock truncations, or alternative squeezing regimes. In such cases, full quantum simulation remains the only dependable evaluation method.
	
	\section{Conclusion}
	\label{sec:conclusion}
	
	We have presented a two-stage gradient-boosted surrogate pipeline for rapid
	screening of GBS circuits targeting GKP state generation.
	Trained on 689 circuit configurations across 3--5 optical modes, the surrogate
	achieves $90.0\%$ GKP-detection accuracy on a held-out set---a $23.7$
	percentage-point improvement over the majority-class baseline---with fidelity
	MAE of $0.032$ and log-scale post-selection $R^2 = 0.837$.
	For the most computationally demanding configurations ($\nmax = 12$, five modes),
	this reduces a ten-thousand-circuit search from ${\approx}12{,}500$ CPU-hours
	to ${\approx}1{,}250$: a tenfold acceleration that makes large-scale parameter
	exploration practical.
	
	The surrogate has clearly defined limits.
	Cross-validation fidelity $R^2 = 0.689 \pm 0.294$ reveals heterogeneous
	reliability, and the $36\%$ Stage~1 misclassification rate compounds into
	Stage~2 errors that aggregate metrics understate.
	Out-of-distribution squeezing conventions cause fidelity errors up to
	$\Delta F = 0.928$, which are caught and contained by the mandatory exact
	simulation step---the essential reliability guarantee that separates the system
	from an autonomous predictor.
	
	Three improvements would close the remaining gaps: symmetry-augmented training
	to resolve sign-convention sensitivity; active learning to reduce
	cross-validation variance; and the addition of Wigner logarithmic negativity as
	a regression target.
	These extensions would move the pipeline toward a fully reliable surrogate for
	autonomous GKP circuit design confirmed by quantum simulation.

	\section*{CRediT authorship contribution statement}
	All authors had an equal and shared role in all stages of the research, including conceptualization, study design, data collection and analysis, writing, and revision of the manuscript.
	
	\section*{Declaration of competing interest}
	The authors confirm that there are no known competing financial interests or personal relationships that could have influenced the work reported in this paper.
	
	\section*{Data availability}
	The data that support the findings of this study are available from the corresponding author upon reasonable request.	
\newpage	
	
	\section*{References}
	\bibliographystyle{unsrt}
	\bibliography{vosq-1}	
	
\end{document}